\documentclass{ws-procs9x6}

\newcommand{\beq}{\begin{equation}}
\newcommand{\eeq}{\end{equation}}
\newcommand{\bqa}{\begin{eqnarray}}
\newcommand{\eqa}{\end{eqnarray}}
\newcommand{\checked}[1]{}

\begin{document}

\title{Progress in Anisotropic Plasma Physics}

\author{Paul Romatschke}
\address{Institut f\"ur Theoretische Physik \\ Technische Universit\"at Wien \\
	Wiedner Hauptstrasse 8-10 \\ A-1040 Vienna, Austria}

\author{Michael Strickland}
\address{Institut f\"ur Theoretische Physik \\ Technische Universit\"at Wien \\
	Wiedner Hauptstrasse 8-10 \\ A-1040 Vienna, Austria}

\maketitle

\abstracts{In 1959 Weibel demonstrated that when a QED plasma has a temperature 
anisotropy there exist unstable transverse magnetic excitations which grow 
exponentially fast.  In this paper we will review how to determine the growth 
rates for these unstable modes in the weak-coupling and ultrarelativistic limits 
in which the collective behavior is describable in terms are so-called ``hard-loops''.  
We will show that in this limit QCD is subject to instabilities which 
are analogous to the Weibel instability in QED.  The presence of such 
instabilities dominates the early time evolution of a highly anisotropic plasma; 
however, at longer times it is expected that these instabilities will saturate 
(condense).  We will discuss how the presence of non-linear interactions between 
the gluons complicates the determination of the saturated state.  In order to 
discuss this we present the generalization of the Braaten-Pisarski isotropic 
hard-thermal-loop effective action to a system with a temperature anisotropy in 
the parton distribution functions.  The resulting hard-loop effective action can 
be used to determine the time and energy scales associated with the possible 
saturation (condensation) of the gluonic modes.  We will also discuss the effects
of anisotropies on observables, in particular on the heavy quark energy loss.}

\section{Introduction}

In the last few years a more or less standard picture of the early stages of a 
relativistic heavy ion collision has emerged.  In its most simplified form 
there are three assumptions: (1) that the system is boost invariant along the 
beam direction, (2) that it is homogenous in the directions perpendicular to the 
beam direction, and (3) that the physics at early times is dominated by gluons 
with momentum at a ``saturation'' scale $Q_s$ which have occupation numbers of 
order $1/\alpha_s$. The first two assumptions are reasonable for describing the 
central rapidity region in relativistic heavy-ion collisions.  The third 
assumption relies on the presence of gluonic ``saturation'' of the nuclear 
wavefunction at very small values of the Bjorken variable $x$ \cite{saturation}. 
In this regime one can determine the growth of the gluon distribution by 
requiring that the cross section for deep inelastic scattering at fixed $Q^2$ 
does not violate unitarity bounds.  As a result the gluon distribution 
function saturates at a scale $Q_s$ changing from $1/k_\perp^2 \rightarrow 
\log(Q_s^2/k_\perp^2)/\alpha_s$.  Luckily, despite this saturation, due to the 
factor of $1/\alpha_s$ in the second scaling relation the occupation number of 
small-$x$ gluonic modes in the nuclear wavefunction is still large enough to 
determine their distribution function analytically using classical nonlinear 
field theory \cite{saturation}.  

These distribution functions are used as input for the subsequent thermalization 
of the quark-gluon plasma.  At short times, $\tau_0 \sim Q_s^{-1}$, the gluon 
distribution function is isotropic, however, after the system begins to expand it 
rapidly develops a large anisotropy between the transverse and beam directions 
with $p_\perp \gg p_z$ due to the fact that initially $p_z \sim \tau^{-1}$.  In 
the weak-coupling limit the assumptions above have been used by Baier et al. in 
an attempt to systematically describe the early stages of quark-gluon plasma 
evolution in a framework called ``bottom-up'' thermalization \cite{BMSS:2001}. 
Using collisional mechanisms they are able to show that during the first stage 
of evolution hard gluons scatter out-of-plane counteracting the effect of the 
expansion of the system reducing the rate at which the longitudinal momentum 
decreases instead to $p_z \sim Q_s^{2/3} \tau^{-1/3}$. Although less extreme in 
terms of the rate at which the longitudinal momenta decreases this scenario 
still results in considerable momentum-space anisotropies in the gluon 
distribution function.  In such anisotropic systems it has been shown that 
the physics of the QCD collective modes changes dramatically compared to 
the isotropic case and instabilities are present which can accelerate the 
thermalization and isotropization of the plasma 
\cite{Weibel:1959,SM,RM:2003,RS:2003,ALM:2003,MRS,RS:2004,AL:2004}.

In addition to their role in isotropization and thermalization it is also 
important to have an understanding of what the impact of such instabilities will 
be on observables at RHIC and LHC energies.  Here we will discuss one test 
observable -- the heavy fermion energy loss  \cite{RS:2004el1,RS:2004el2}. One 
might be worried that there is a fundamental problem with perturbation theory 
since the presence of instabilities naively causes the calculation of the soft 
contribution to the heavy quark energy loss to be divergent; however, there is 
protection mechanism dubbed ``dynamical shielding'' which renders the 
collisional energy loss finite for QED and QCD \cite{RS:2004el1,RS:2004el2}. 
However, at realistic couplings the presence of instabilities and associated 
poles on neighboring Riemann sheets \cite{RS:2004} causes significant changes in 
the soft energy loss contribution for both QED and QCD. In fact, as we will 
discuss, the unphysical poles can even change the sign of the heavy quark energy 
loss at low momentum turning it instead into energy \emph{gain}.

\section{Anisotropic Gluon Polarization Tensor}

We consider a quark-gluon plasma with a parton distribution function
which is decomposed as
\bqa
f({\bf p}) \equiv 2 N_f \left(n({\bf p}) + \bar n ({\bf p})\right) + 4 N_c n_g({\bf p}) \; ,
\label{distfncs}
\eqa
\checked{mp}
where $n$, $\bar n$, and $n_g$ are the distribution functions of quarks, anti-quarks, 
and gluons, respectively, and the numerical coefficients collect all appropriate
symmetry factors.
Using the result of Ref.~[\refcite{RS:2003}] the spacelike 
components of the high-temperature gluon self-energy for gluons with soft 
momentum ($k \sim g T$) can be written as
\begin{equation}
\Pi^{i j}(K) = - \frac{g^2}{2} \int \frac{d^3{\bf p}}{(2\pi)^3} v^{i} \partial^{l} f({\bf p})
\left( \delta^{j l}+\frac{v^{j} k^{l}}{K\cdot V + i \epsilon}\right) \; ,
\label{selfenergy2}
\end{equation}
\checked{mp}
where the parton distribution function $f({\bf p})$ is, in principle, 
completely arbitrary.  In what follows we will assume that $f({\bf p})$ can be 
obtained from an isotropic distribution function by the rescaling of only one 
direction in momentum space.  In practice this means that, given any isotropic 
parton distribution function $f_{\rm iso}(p)$, we can construct an 
anisotropic version by changing the argument of the isotropic distribution 
function
\begin{equation}
f({\bf p}) = \sqrt{1+\xi} \, f_{\rm iso}\left(\sqrt{{\bf p}^2+\xi({\bf p}\cdot{\bf \hat n})^2}\right) \; ,
\label{distfunc2}
\end{equation}
\checked{mp}
where the factor of $\sqrt{1+\xi}$ is a normalization constant which ensures 
that the same parton density is achieved regardless of the anisotropy introduced, 
${\bf \hat n}$ is the direction of the anisotropy, and $\xi>-1$ is an adjustable 
anisotropy parameter with $\xi=0$ corresponding to the isotropic case. 
Here we will concentrate on $\xi>0$ which corresponds to a 
contraction of the distribution along the ${\bf \hat n}$ direction since this is 
the configuration relevant for heavy-ion collisions at early times, namely
two hot transverse directions and one cold longitudinal direction.

Making a change of variables 
in (\ref{selfenergy2}) it is possible to integrate out the $|p|$-dependence 
giving~\cite{RS:2003}
\begin{equation}
\Pi^{i j}(\omega/k,\theta_n) = \mu^2 \int \frac{d \Omega}{4 \pi} v^{i}%
\frac{v^{l}+\xi({\bf v}\cdot\hat{\bf n}) \hat{n}^{l}}{%
(1+\xi({\bf v}\cdot\hat{\bf n})^2)^2}
\left( \delta^{j l}+\frac{v^{j} k^{l}}{K\cdot V + i \epsilon}\right) ,
\end{equation}
\checked{mp}
where $\cos\theta_n \equiv \hat{\bf k}\cdot\hat{\bf n}$ and $\mu^2 \equiv 
\sqrt{1+\xi}\;m_D^2>0$. The isotropic Debye mass, $m_D$, depends on 
$f_{\rm iso}$.  In the case of pure-gauge QCD with an equilibrium 
$f_{\rm iso}$ we have $m_D = g T$.

The next task is to construct a tensor basis for the 
spacelike components of the gluon self-energy and propagator.  We therefore need 
a basis for symmetric 3-tensors which depend on a fixed anisotropy 3-vector 
$\hat{n}^{i}$ with $\hat{n}^2=1$.  This can be achieved with the following four component 
tensor basis: $A^{ij} = \delta^{ij}-k^{i}k^{j}/k^2$, $B^{ij} = k^{i}k^{j}/k^2$, 
$C^{ij} = \tilde{n}^{i} \tilde{n}^{j} / \tilde{n}^2$, and $D^{ij} = 
k^{i}\tilde{n}^{j}+k^{j}\tilde{n}^{i}$ with $\tilde{n}^{i}\equiv A^{ij} \hat{n}^{j}$. 
Using this basis we can decompose the self-energy into four structure functions 
$\alpha$, $\beta$, $\gamma$, and $\delta$ as ${\bf \Pi}= \alpha\,{\bf A} + 
\beta\,{\bf B} + \gamma\,{\bf C} + \delta\,{\bf D}$.~\footnote{ Explicit analytic 
integral expressions for $\alpha$, $\beta$, $\gamma$, and $\delta$ can be found 
in Ref.~[\refcite{RS:2003}].  Additionally, analytic expressions in the small-$\xi$
limit and for propagation along the anisotropy direction can be found in 
Refs.~[\refcite{RS:2003}] and [\refcite{RS:2004}], respectively.}

\section{Collective Modes}

As shown in Ref.~[\refcite{RS:2003}] this tensor basis allows us to express the 
propagator in terms of the following three functions
\bqa
\Delta_\alpha^{-1}(K) &=& k^2 - \omega^2 + \alpha \; , \label{propfnc1} \nonumber \\
\Delta_\pm^{-1}(K) &=& \omega^2 - \Omega_\pm^2 \; , \nonumber
\label{propfnc2}
\eqa
where $ 2 \Omega_{\pm}^2 = \bar\Omega^2 \pm \sqrt{\bar\Omega^4- 4 
((\alpha+\gamma+k^2)\beta-k^2\tilde n^2\delta^2) }$
and $\bar\Omega^2 = \alpha+\beta+\gamma+k^2$.  

Taking the static limit of these 
three propagators we find that there are three mass scales:  $m_\pm$ and $m_\alpha$. In 
the isotropic limit, $\xi\rightarrow0$, $m_\alpha^2=m_-^2=0$ and $m_+^2 = m_D^2$.  
However, for $\xi>0$ we find that $m_\alpha^2<0$ for 
all $\mid\!\!\theta_n\!\!\mid\,\neq\pi/2$ and $m_-^2<0$ for 
all $\mid\!\!\theta_n\!\!\mid\,\leq\pi/4$.  Note also that for $\xi>0$ both $m_\alpha^2$ and $m_-^2$
have there largest negative values at $\theta_n=0$ where they are equal.

The fact that for $\xi>0$ both $m_\alpha^2$ and $m_-^2$ can be negative 
indicates that the system is unstable to both magnetic and electric fluctuations 
with the fastest growing modes focused along the beamline ($\theta_n=0$).   In 
fact it can be shown that there are two purely imaginary solutions to each of the 
dispersions relations $\Delta_\alpha^{-1}(K)=0$ and $\Delta_- ^{-1}(K)=0$ with the 
solutions in the upper half plane corresponding to unstable modes.  We can 
determine the growth rate for these unstable modes by taking $w\rightarrow 
i\Gamma$ and then solving the resulting dispersion relations for $\Gamma(k)$.

\begin{figure}
\begin{minipage}[t]{4.5cm}
\vspace{-5.1cm}
\hspace{6mm}In Fig.~1 we plot the instability growth rates, $\Gamma_\alpha$ and $\Gamma_-$, as a function
 of wave number for $\xi=10$ and $\theta_n=\pi/8$.  Note that both growth rates vanish at $k=0$ and
 have a maximum $\Gamma_*\sim\mu/10$ at $k_*\sim\mu/3$.  The fact that they have a maximum
 means that at early times the system will be dominated by unstable modes with spatial frequency
 $1/k_*$.
 
\end{minipage}
$\;\;\;$
\begin{minipage}[t]{6.2cm}
\includegraphics[width=6.2cm]{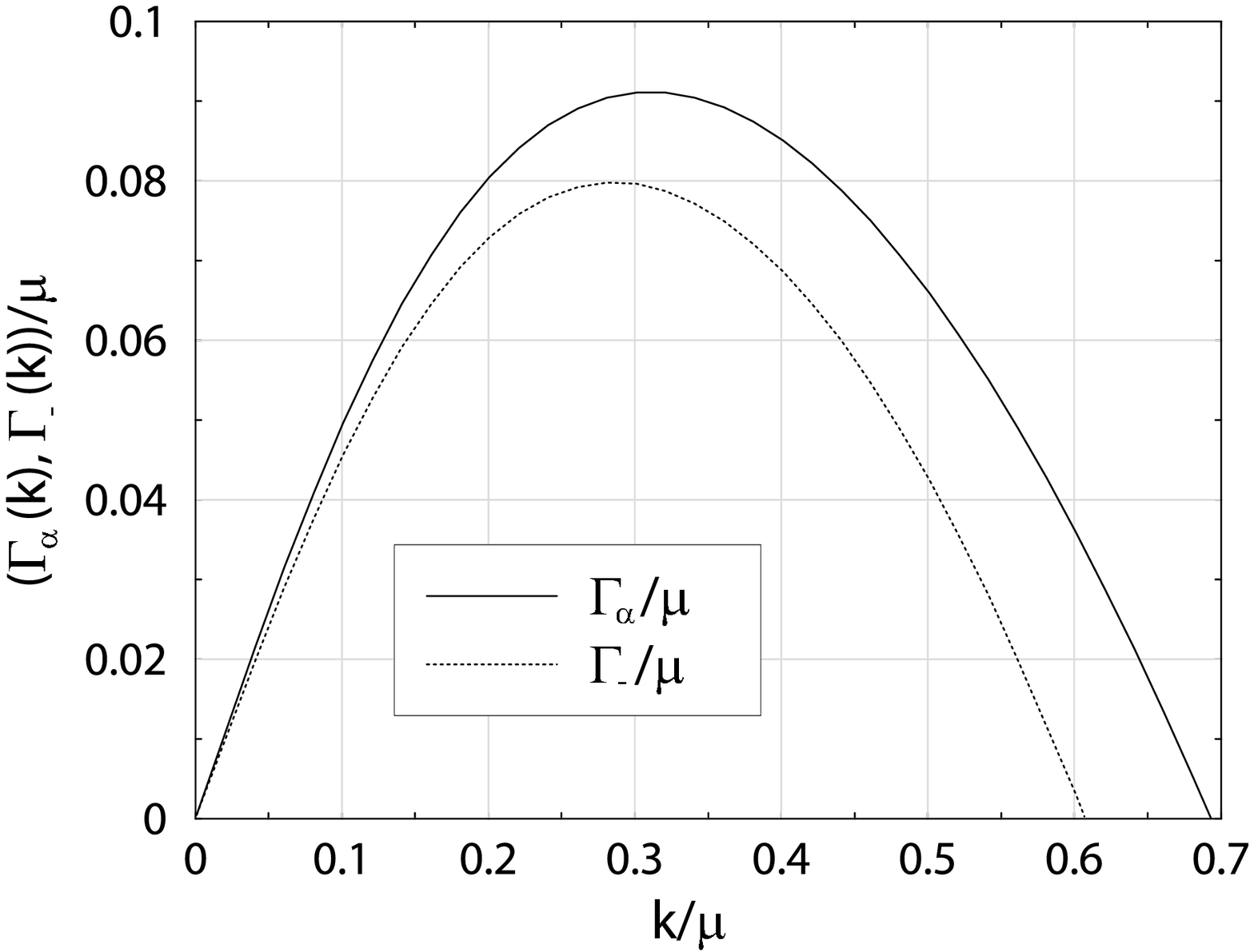}
\begin{center}
\vspace{-2mm}
{\footnotesize $\;\;\;\;\;$ Fig.~1: Instability growth rates. }
\end{center}
\end{minipage}
\label{fig1}
\end{figure}
\vspace{-8mm}

\section{The Hard-Loop Effective Action}

In the previous section by only considering the polarization tensor we 
implicitly used a linear analysis which suffices only during very early times. 
As the instabilities grow non-linear interactions become more important and can 
halt their growth.  In QED simulations such a saturated state does emerge and is called a 
Bernstein-Greene-Kruskal wave; however, in QCD 
any corresponding saturated state would necessarily be a much more complex 
beast.  Answering these questions requires knowledge of the full hard-loop 
effective Lagrangian in an anisotropic system.  Luckily, it is possible to 
obtain a simple expression which generates all hard-loop (HL) vertex functions 
\cite{MRS}
\bqa
{\mathcal L} &=& -\frac{1}{4} F_{\mu\nu} F^{\mu\nu}
+\frac{g^2}{2} \int_{\mathbf p}
\frac{f(\mathbf p)}{|{\bf p}|}
F_{\mu \nu}(x)
\frac{p^\rho p^\nu}{(p \cdot D)^2} \;
F_\rho^{\;\;\mu} (x)
\label{lag}
\eqa
For example, from this we can obtain the hard-loop
3-gluon vertex
\bqa
\Gamma^{\mu\nu\lambda}_{\rm HL}(k,q,r) &=& 
\frac{g^2}{2}
\int_{\bf p} 
\frac{ \partial f({\bf p})}{\partial p^\beta } 
\;
\hat{p}^\mu \hat{p}^\nu \hat{p}^\lambda
\left( 
\frac{r^\beta}{\hat{p}\!\cdot\!q\;\hat{p}\!\cdot\!r}
-\frac{k^\beta}{\hat{p}\!\cdot\!k\;\hat{p}\!\cdot\!q}
\right) \; .
\label{3gvertex}
\eqa
In contrast to the isotropic $n$-gluon HL vertices which vanish in the static 
limit, the anisotropic $n$-gluon HL vertices can be non-zero.  This opens up the 
possibility for additional saturation mechanisms which are governed by soft non-abelian 
physics sensitive to these higher vertices.

Note that Arnold and Lenaghan have shown that if one considers fluctuations in 
the vector potential which are directed along the anisotropy direction and only 
depend on the longitudinal coordinate then it is possible to reduce the 
effective lagrangian (\ref{lag}) to a quadratic form which simplifies the 
analysis considerably \cite{AL:2004}.  They then make some additional 
assumptions including ignoring time-depedence in the self-energy in order to 
construct a toy-model which is then used to study the subsequent evolution of 
instabilities in the system.  Their results indicate that in non-abelian gauge 
theories that possible non-abelian saturated states are metastable and as a 
result instability growth forces the system along abelian directions. However, 
it should be noted that by ignoring the time dependence in the potential as they 
have done, the instability growth rate in their model is finite at $k=0$. This 
is not the case when time dependence is included in the full analysis and 
instead one finds that the growth rate at $k=0$ then vanishes as shown in Fig. 
1. This difference could have a significant impact on the abelianization 
particularly on the ``abelianization length'' measured in their simulations.

\section{Effects on observables}

Since the unstable modes manifest themselves as poles of
the propagator in the static limit, it has been argued \cite{MooreBiel} 
that the presence of these instabilities in general prohibits
the calculation of observables in a perturbative framework, since 
those quantities would be plagued by unregulated divergences.

However, it turns out that at least one observable, namely the collisional
energy loss of a heavy fermion, is protected from these unregulated 
divergencies by a mechanism dubbed ``dynamical shielding'' \cite{RS:2004el1}
(see Ref.~[\refcite{Rdiss}] for a proof in the case of $\xi\ll 1$ and 
$\xi\rightarrow \infty$). As a consequence of dynamical shielding
(which incidently is somewhat similar to dynamical screening of the magnetic
interaction in QCD, hence the name), the collisional energy loss
becomes calculable perturbatively also in the anisotropic case. Therefore,
we will use the collisional energy loss as ``test observable'' 
to learn something about the effects
of the instabilities on physical quantities. 

Schematically, the soft contribution to the collisional energy loss is given by
\beq
-\left(\frac{{\rm d} W}{{\rm d} t}\right)_{\rm soft} \sim 
  \, {\rm Im} \, \int \frac{d^3 {\bf k}}{(2 \pi)^3} \, k \, z \,
\Delta(z,k,\theta_n)  \; ,
\label{elosssimple}
\eeq
where $\Delta(z,k,\theta_n)=v^i \Delta^{ij} v^j$ 
is the relevant contraction of the propagator 
\cite{RS:2004el1} and $z={\bf \hat{k}}\cdot {\bf v}$ with ${\bf v}$ 
the velocity of
the heavy fermion. The instabilities would in principle affect the
integrand for ${\rm Re}\, z=0$ causing singularties to be encountered
along the momentum integration path.  In order to see that there are,
in fact, no singularities we focus on terms of the form 
$(q^2 + \alpha)^{-1}$.  
Due to the fact that in the static limit the 
structure function $\alpha$ is negative-valued we could encounter a
singulartity at $q^2=\alpha$; however, if one takes the static 
limit of $\alpha$ carefully we find
\bqa
\lim_{z\rightarrow0} \alpha(z) = M^2(-1 + i D z)  + {\mathcal O}(z^2) \; ,
\eqa
where $M$ and $D$ depend on the angle of propagation with respect to the 
anisotropy vector and the strength of the anisotropy.  As long as $D$ is 
non-vanishing the singularities are regularized because of the $z$ in the numerator 
of Eq.~(\ref{elosssimple}) and we call the singularity ``dynamically shielded''. 
This effect can be shown to regulate all singularities which would have naively
come from the presence of instabilities.
Therefore, there is no catastrophic effect from the instabilities on the 
collisional energy loss. However, there is an effect which is associated with 
the presence of the instabilities; in order to understand this effect it is 
however necessary to extend the analysis of the collective modes of an 
anisotropic plasma to arbitrary Riemann sheets.

\begin{figure}[t]
\begin{center}
\begin{center}
\includegraphics[width=0.5\linewidth]{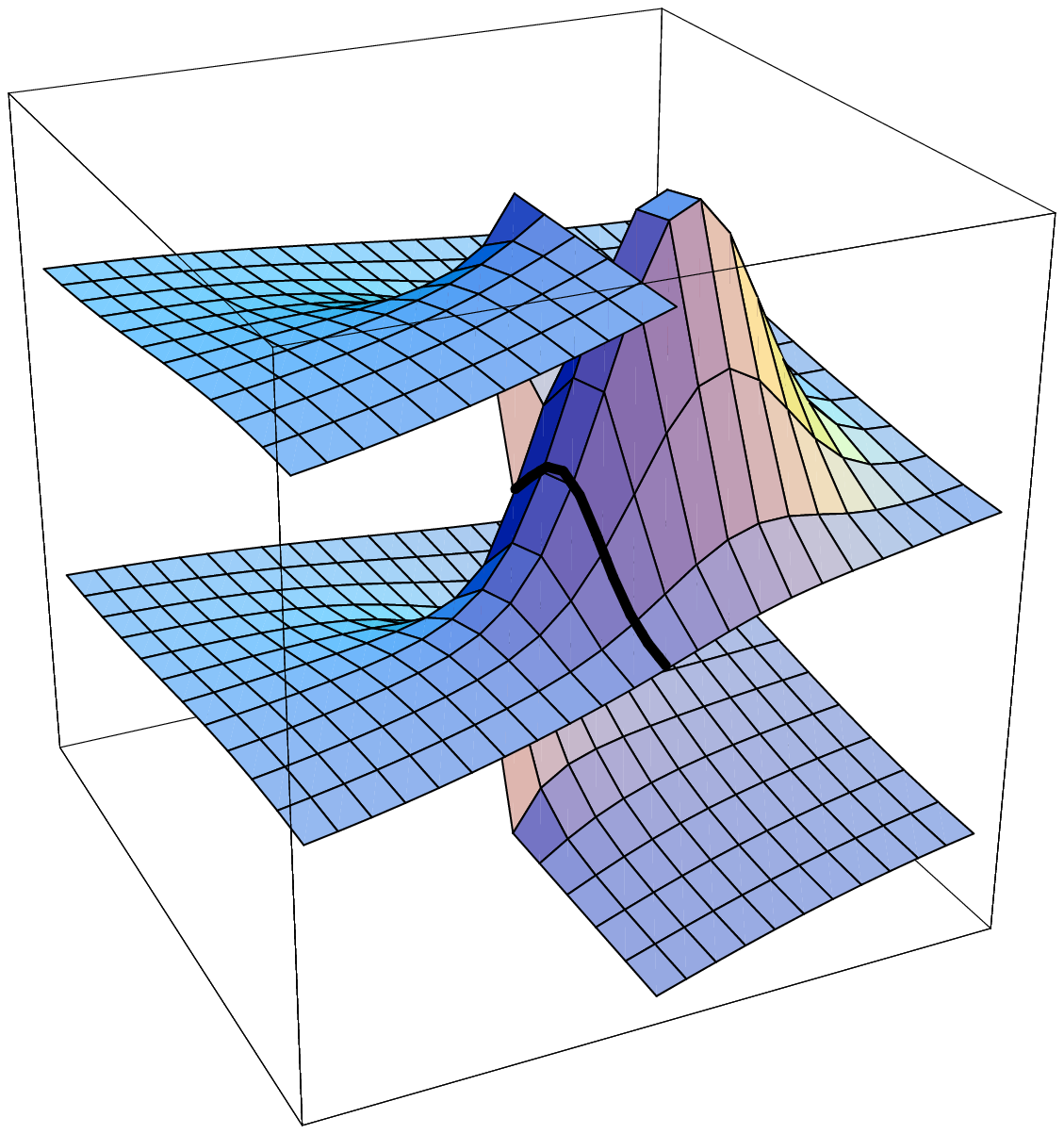}
\setlength{\unitlength}{1cm}
\begin{picture}(9,0)
\put(1.8,1.5){\makebox(0,0){\footnotesize ${\rm Re}\,z$}}
\put(5.5,0.6){\makebox(0,0){\footnotesize ${\rm Im}\,z$}}
\end{picture}
\vbox{
\parbox[t]{10cm}{\footnotesize
Fig.~2: Sketch of the complex $z$ plane including the extension
of the logarithm to the unphysical sheet. Also shown
is how a pole in the unphysical region (mountain) has effects felt on the
physical sheet.  The black line indicates where the two sheets are joined
together.
}
}
\end{center}
\label{fig2}
\end{center}
\end{figure}

The extension to the unphysical Riemann sheets is achieved by continuing
analytically the structure functions beyond their cut structures (which
for finite $\xi$ is a logarithmic cut running along the real $z$ axis for
$z^2<1$, in complete analogy to the isotropic case, but a square-root cut
for $z^2<\sin^2{\theta_n}$ for $\xi\rightarrow \infty$)\cite{RS:2004}.
More precisely, since one can continue the structure functions from
either below or above the cut, the physical sheet has two neighboring
unphysical Riemann sheets, which -- at least in the case of finite $\xi$ --
could themselves be extended to reproduce the well-known ``spiral staircase''
form of the complex logarithm. 
Once the structure functions on the unphysical sheets are known, one may 
conduct the analysis of the collective modes as was done
on the physical sheet, finding -- among others -- singularities for
complex $z$ on the neighboring unphysical sheets that extend to the region
of spacelike momenta~\cite{RS:2004}.

But why bother about these modes, given that they do not ``live'' on the 
physical sheet? To see that there can be an effect on physical quantities, 
imagine once again the spiral staircase spanned by the complex logarithm: the 
physical sheet would correspond to the region covered by the spiral plane from 
the ground floor to the first floor, while the unphysical sheet where the extra 
quasiparticle mode lives would correspond to the region first to second floor. 
Since the existence of such a mode corresponds to a pole in the propagator, 
there is has a mountainous dent (singularity) in the spiral plane somewhere from 
the first to the second floor.  However, since the mountain has a finite width, 
its base can be felt also below the second floor, especially if its peak is near 
the first floor (see Fig.~2 for a sketch). Therefore, the nearer a mode on a 
neighboring unphysical sheet comes to the border of the physical sheet, the more 
will it affect the propagator on the physical sheet and consequently any 
physical observable sensitive to the relevant region of phase-space (in this 
case spacelike momentum).

This is precisely the situation encountered for the collisional energy loss: 
there are unstable modes on the physical sheet which, once the momentum $k$ 
becomes larger than some critical value (see Fig.~1), move onto the neighboring 
unphysical sheets and subsequently influence the propagator on the physical 
sheet by the mechanism explained above. In fact, it turns out that the the 
contributions from these unphysical modes may drive the leading-order 
perturbative results for the collisional energy loss \emph{negative} for small 
fermion velocities, corresponding to an energy transfer from hard to soft 
scales. Note, however, that this negative energy loss is \emph{not} due to the heavy 
fermion in question having a sub-thermal velocity but is instead connected to 
the instabilities of an anisotropic system, as already noted in 
Ref.~[\refcite{Lif81}].

For further discussion of the heavy quark energy loss in an anisotropic quark-gluon 
plasma we refer the reader to Ref.~[\refcite{RS:2004el2}].  In addition to 
elucidating the impact of unstable modes and associated poles on unphysical 
Riemann sheets, in that paper we showed that for anisotropic systems the heavy 
quark collisional energy loss has an angular dependence which increases as the 
coupling and/or anisotropy is increased. Quantitatively, for $\alpha_s=0.3$ and 
a 20 GeV bottom quark we found that the deviations from the isotropic result 
were on the order of 10\% for $\xi=1$ and of the order of 20\% for $\xi \geq 
10$.  When translated into the difference between longitudinal and transverse 
energy loss this results in a 10\% difference at $\xi=1$, a 30\% difference at 
$\xi=10$, and a 50\% difference at $\xi=\infty$.  

\section{Conclusions}

In this paper we have attempted to summarize the developments which have occured 
in the last two years regarding the physics of a quark-gluon plasma which has a 
momentum-space anisotropy in the underlying parton distribution functions.  We 
have presented expressions for the self-energy and full effective action in this 
case and discussed the emergence of unstable modes which cause exponential 
growth of the gluon field along the anisotropy direction. In addition, we 
discussed the expected effects such anisotropies have on observables 
concentrating on the heavy quark energy loss as a test observable.

As we have tried to illustrate here the physics of anisotropic plasmas is 
qualitatively different from isotropic ones.  In the weak-coupling limit the 
short-time behaviour of anisotropic plasmas is dominated by the growth of 
unstable modes.  In order to properly understand thermalization and 
isotropization of a quark-gluon plasma it seems necessary to take into account 
the effect of such unstable modes from the earliest times.  We note in closing 
if a Vlasov description is applicable at very early times ($\tau_0 
\sim Q_s^{-1}$) then instability broadening of the longitudinal momentum is much 
more effective than collisional broadening and it is possible that anisotropies 
due to expansion of the system are, in fact, never generated or reduced 
significantly. However, a more conservative approach calls for the Vlasov 
description to only be applied once the hard gluon occupation number becomes 
less than one which according to collisional arguments occurs at a time $\tau_1 
\sim \alpha^{- 3/2} Q_s^{-1}$. If this more conservative approach is applicable 
then the anisotropy at $\tau_1$ can be parametrically estimated to be $\xi \sim 
\alpha_s^{-3/2}$ ($\xi\sim\alpha_s^{-1/2}$ if collisional broadening is taken in 
account) which results in a strongly anisotropic system in the weak-coupling 
limit. Whichever approach is used, however, it is clear that it is necessary to 
take into account the unstable modes.

\section*{Acknowledgements}

The authors would like to thank A.~Rebhan and the organizers of the SEWM
conference.  In addition, P.R. would like to thank the University of Helsinki
for financial support and B.~Heher for his support and guidance.

\end{document}